%% file: ferrigno_1E1145.tex
\documentclass[]{aa}

\usepackage{amsmath}
\usepackage[dvips]{graphicx}
\usepackage{natbib}
\bibpunct{(}{)}{;}{a}{}{,}
\usepackage[american]{babel}

\input{mymacros.tex}

\input{aasm.tex}

\begin{document}

\title{\textsl{INTEGRAL} observation of the accreting pulsar 1E1145.1-6141}

\author{
Carlo Ferrigno
\inst{1}
\and
Alberto Segreto\inst{1} 
\and
Teresa Mineo\inst{1}
\and
Andrea Santangelo\inst{2}
\and
R\"udiger Staubert\inst{2}
}

\authorrunning{C. Ferrigno et al}
\titlerunning{\textsl{INTEGRAL} observation of 1E~1145.1-6141}
\offprints{C. Ferrigno, \email{ferrigno@ifc.inaf.it}}

\institute{IASF-Pa, via Ugo la Malfa 153, 90146 Palermo Italy
    \and
    IAAT, Abt.\ Astronomie, Universit\"at T\"ubingen,
    Sand 1, 72076 T\"ubingen, Germany}

\date{Received \today; accepted ---}

\abstract
   {}
   {
We analyze 1050\,ks of \textsl{INTEGRAL} data of the high mass X-ray
binary pulsar 1E~1145.1-6141 to study its 
properties over a long time baseline, from June 2003 to June 2004,
with wide spectral coverage.
}
{
We study three high luminosity episodes, two of them at
the system apoastron, three brightening with lower intensity, two at
the periastron, and one extended period of intermediate luminosity
spanning one orbital cycle.
We perform timing analysis to determine the pulse period and
pulse profiles at different energy ranges. We also analyze the
broad band phase average spectrum of different luminosity states
and perform phase resolved spectroscopy for the first
flare.
}
{
From the timing analysis, we find a pulse period of $\sim297$\,s
around MJD~53\,000 with a significant scatter around the mean value.
From the spectral analysis
we find that the source emission can be described by an absorbed
bremsstrahlung model in which the electron temperature varies between
$\sim$25 and $\sim$37\,keV, without any correlation to luminosity, and the
intrinsic absorbing column is constantly of the order of
$10^{23}\,\mathrm{cm^{-2}}$.
Phase resolved spectral analysis evidences a
different temperature of the plasma
in the ascending and descending edges of the pulse during the first flare.
This justifies the pulse maximum shift by $\sim0.4$ phase units between 20 and
100\,keV observed in the pulse profiles.
}
{
The comparison with the previous period measurements
reveals that the source is currently spinning-down,
in contrast to the long term secular trend observed so far
indicating that at least a temporary accretion disk is formed. The
study of the spectral property variations with respect to time and
spin phase suggests the presence of two emitting components at
different temperatures whose relative intensity varies with time.
}
{}


\maketitle

\section{Introduction}

X--ray Binary Pulsars (XRBPs) were discovered more than 30 years ago
with the pioneer observation of Cen~X--3 by \citet{giac71}.
Although the basic mechanisms of the pulsed emission were
quickly understood \citep{Pringle72,Davids73}, many puzzling aspects still persist.
In particular,
their wide band spectral behavior has not been completely explained on the
basis of a self consistent physical model; spectra are
successfully described by phenomenological models such as a power law modified
at high-energy by an exponential cut-off and
at low energy by intrinsic photo-electric absorption plus a
thermal component at $\sim0.1$\,keV \citep{coburn2002,disalvo2004}.
XRBPs are powered by the accretion of ionized gas
from a nearby companion, an O or B star ($M \geq 5M_{\odot}$) in
High Mass X-ray Binaries (HMXBs), a later than type A star
or a white dwarf ($M\leq M_{\odot}$) in Low Mass X-ray Binaries (LMXBs)
onto the surface of a black hole or neutron star (NS).
If the compact object is highly magnetized, with surface field
as high as $10^{12}$\,G,
the plasma is threaded at the magnetospheric boundary, 
and then funneled onto the magnetic poles \citep{Pringle72,Davids73} along the field lines;
pulsation is generated because of the
nonalignment between the magnetic and the rotational axes.
In HMXBs the plasma can be captured directly from the slow and dense equatorial stellar wind
of a Be star, or from the radiation driven wind of an early type super-giant.
In brighter systems, accretion can be also powered
by Roche-lobe overflow via an accretion disk.

The systems with Be companions (often with appreciable orbital eccentricity)
are normally quiescent but sometimes they flare when
the NS passes through the companion's disk.
During these episodes the luminosity may increase
by a large factor, a temporary accretion disk may form, and the NS generally
spins up.
The bright HMXBs accreting from a stable accretion disk via Roche lobe overflow have rather short spin
and orbital periods, nearly circular orbits, and can exhibit torque reversal episodes superimposed on
a spin-up trend as observed for Cen~X--3 \citep{bildsten1997}.
Most wind-fed HMXBs with a super-giant companion, 
have long pulse periods (hundreds of seconds), 
moderate and rather stable luminosity
($L_X \simeq 10^{35-37}\,\mathrm{erg\,s^{-1}}$), and can
spin-up or spin-down depending on the transfer of angular momentum 
from the wind to the magnetosphere.
The pulse period evolution is then characterized by red-noise fluctuations due to the irregular accretion flow 
as in the case of Vela~X--1 \citep{bildsten1997}.
Simulations have shown that a prograde or retrograde accretion disk,
which is not necessarily a true viscous disk,
may form and disrupt on time scales of $10^3$\,s, causing changes in the sign
of the pulse period derivative \citep{blondin1990}, as observed in
the controversial case of the system with unknown companion OAO~1657-415
\citep{baykal1997,turchi2000}. A temporary accretion disk is also suggested to explain 
the short episodes of spin-up, accompanied by luminosity enhancements, in the wind-fed binary GX~301--2 
\citep{koh1997,turchi2000}.

Thanks to the Galactic plane monitoring,
performed with the \textsl{INTEGRAL} satellite, a new class of HMXBs containing a neutron star and 
super-giant (SG) donors has been identified: the Super-giant Fast X--ray Transients (SFXT),
characterized by short outbursts of X-ray emission
\citep{negueruela2006,sguera2005,sguera2006}. The transient
behavior of these sources is explained by the motion in wide orbits (~10 stellar radii) 
through a clumpy SG wind \citep{jean2005,negueruela2006},
or to the recurrent passage of the NS through a denser equatorial wind component \citep{sidoli2007}.

In this paper, we report on the \textsl{INTEGRAL} observations of one of the
less known HMXBs located very close ($17^\prime$)
to the extensively studied 4U~1145--61 \citep[see e.g. ][]{stevens1997}, 1E~1145.1--6141.
Optical spectroscopy of its B2~Iae super-giant companion yields a distance
of $(8.5\pm1.5)$\,kpc \citep{densham1982}. X-ray
spectral analysis was first performed using the data in the energy range 0.5--3\,keV 
from the imaging proportional counter of the \textsl{Einstein}
observatory \citep{lamb1980} finding a column density of
$N_\mathrm{H}=(3\pm2)\times10^{22}\,\mathrm{cm}^{-2}$. This value
was confirmed, using the proportional counter array of the \textsl{RXTE} satellite in the energy range 2--60\,keV, 
by \citet{ray2002} who
modeled the spectrum using an absorbed cut-off power law with index
$\Gamma=1.4$, cut-off energy $E_c=6.4$\,keV and $e$-folding energy
$E_f=18$\,keV.
They also measured the orbital parameters of the binary system
finding an eccentric ($e=0.20$)
orbit with period $P=14.365$\,days, $a\sin i = 99.4$\,lt-sec,
periastron epoch $T_0=51008.1$\,MJD, and longitude of periastron
$\omega=-52\gra$. They proposed, on the basis of previous
observations, a secular spin-up trend of $\sim10^{-9}$\,s/s between 1975 and 2000
\citep{ray2002}. The NS emission was characterized by a pulsed fraction of
$\sim$50\% in the 4--20\,keV energy band and an X-ray luminosity of
$\sim10^{36}$\,erg/s \citep{hutchings1981, grebenev1992,ray2002}.
Such a low luminosity is inconsistent with Roche lobe overflow 
for this class of binaries
and indicates that the NS is almost certainly
accreting from the companion wind. The pulse shape has been studied firstly
using the ART-B telescope on board of the \textsl{GRANAT} observatory in the 3--60\,keV energy range
by \citet{grebenev1992} who found evidence of a notch
between the maximum and the minimum of the pulse; this feature
appeared also in the pulse shape obtained with \textsl{RXTE}
\citep{ray2002}.

Our paper is aimed to study the spectral and timing properties of
1E~1145.1--6141 between March 2003 and June 2004 using the data from
European Space Agency's International Gamma-Ray Astronomy Laboratory
(\textsl{INTEGRAL}) and is organized as follows: in
Sect.~\ref{sec:observation} we describe the observations, in
Sect.~\ref{sec:results} we illustrate the scientific results which
are discussed in Sect.~\ref{sec:discussion}, and in
Sect.~\ref{sec:conclusions} we draw our conclusions.

\section{Observation and data analysis}
\label{sec:observation}
\textsl{INTEGRAL}, launched in October 2002, carries 3 co-aligned
coded mask telescopes.
\begin{itemize}
\item \textsl{IBIS} \citep[Imager on Board the
  \textsl{INTEGRAL} Satellite;][]{ibis}, which allows for
12\,arcmin FWHM angular resolution imaging in the energy range from 15\,keV to
600\,keV with energy resolution of 8\% at 100\,keV and
$8\gra$ fully coded field of view. It is composed by a low energy
CdTe detector (15--600\,keV), \textsl{ISGRI} \citep{isgri}, and by a
CsI layer (175\,keV--10\,MeV), \textsl{PICsIT} \citep{picsit}.
\item \textsl{SPI}
\citep[SPectrometer on \textsl{INTEGRAL};][]{spi},
sensitive from 20\,keV to 8\,MeV with an angular resolution
of $2.5\gra$ and an energy resolution of a
0.2\% at 1.3\,MeV; the fully coded field of view is $16\gra$.
\item \textsl{JEM-X} \citep[Joint European
X-ray Monitor;][]{jemx} that includes two independent units \textsl{JEM-X1}
and \textsl{JEM-X2}, sensitive from 3\,keV to 34\,keV, with an
angular resolution of 1\,arcmin and an energy resolution of 13\%
at 10\,keV; its fully coded field of view is $4.8\gra$.
\end{itemize}

\textsl{INTEGRAL} observations are divided in short periods
(of  $\sim2000$\,s each), called science windows (SCWs), during which
the telescope maintains the same pointing. From one science window to another,
the pointing follows a dithering or a survey strategy as chosen in the
observing program \citep{integral}.

To reduce and analyze \textsl{JEM-X} data we used the Off-line Science
Analysis (\texttt{OSA}) software version 5.1;
for \textsl{ISGRI} data we exploited also the software described
in \citet{ferrigno2006} and \citet{segreto2007}. For the spectral analysis we used the
standard \texttt{XSPEC} package version 12.2.

In this paper we used the publicly available
\textsl{INTEGRAL} data in which 1E~1145.1-6141 was visible before
June 2004. Thanks to the ISGRI large field of view it has been
possible to observe the source for a total of $1.08$\,Ms 
between June 2003 and June 2004, with a nearly continuous coverage of 2 binary orbits in 2004,
and a sparse coverage of two and a half orbits in 2003.
We limited our analysis to \textsl{ISGRI} and \textsl{JEM-X} data because the source is too weak
to be studied with \textsl{SPI}. As summarized in Table~\ref{tab:obs},
\textsl{JEM-X} detected the source only in a sub-set of the available SCWs,
because of its smaller field of view with respect to \textsl{ISGRI}.

\begin{figure}
  \begin{center}
     \resizebox{\hsize}{!}{\includegraphics[angle=0,width=6cm]
      {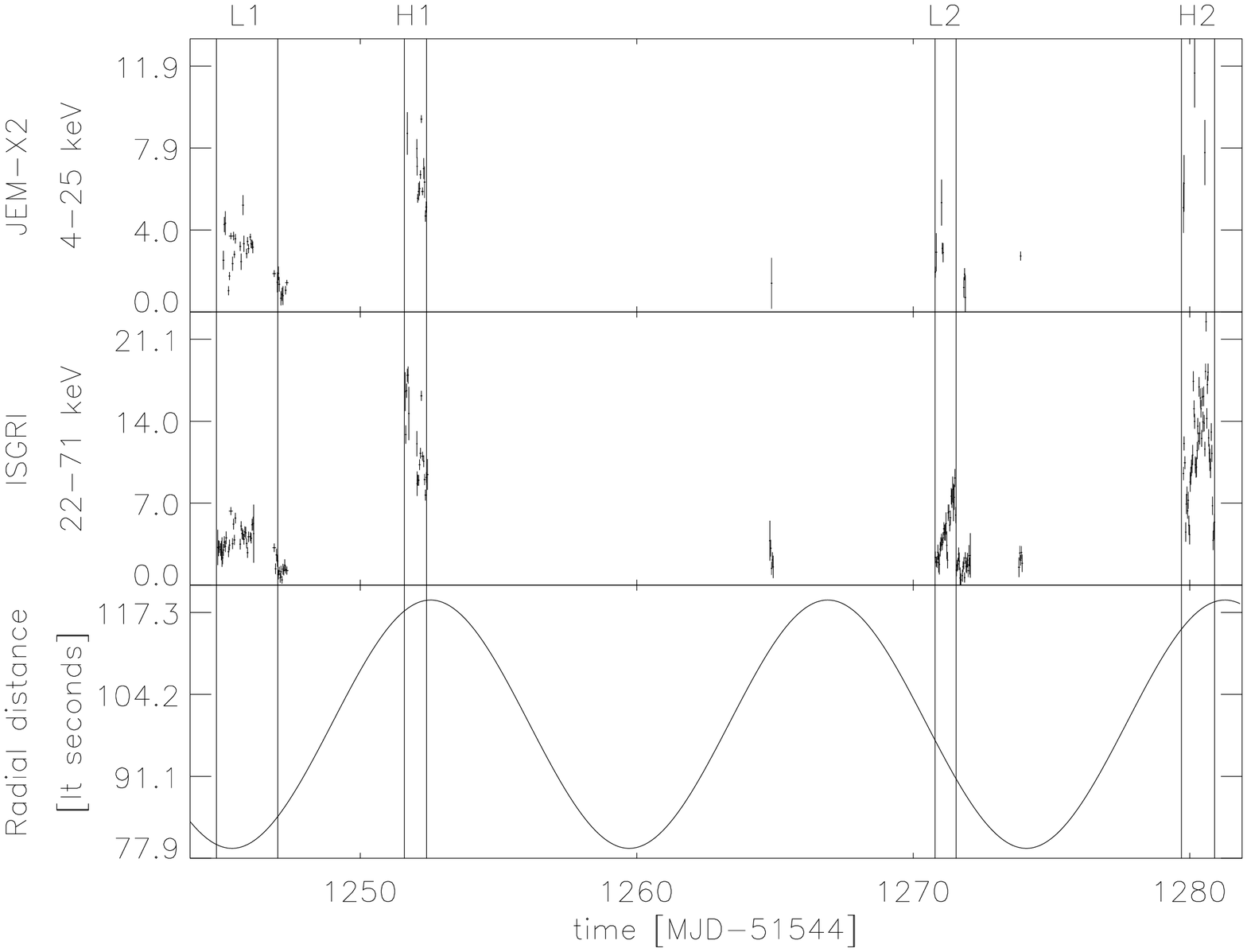}}
     \resizebox{\hsize}{!}{\includegraphics[angle=0,width=6cm]{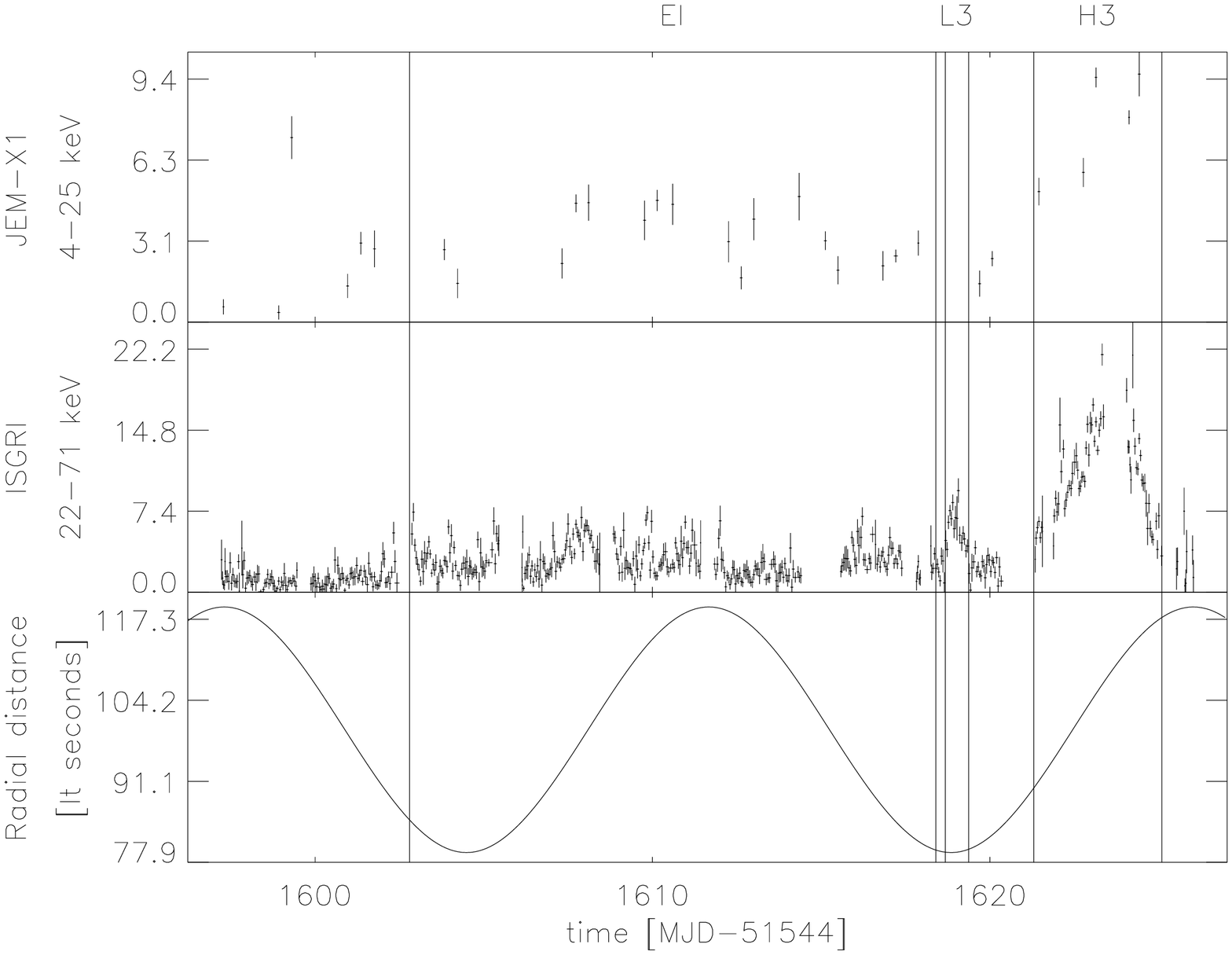}}
    \caption{
    Light curve of 1E~1145-6141 during the \textsl{INTEGRAL}
    observations. In each panel we plot, from top to bottom, the data of
    \textsl{JEM-X} (4--25\,keV) and \textsl{ISGRI} (22--71\,keV),
    (from \textbf{\texttt{http://isdc.unige.ch/?Data+sources}})
    and the distance between the stars of the binary system expressed in light
    seconds 
    as computed from the ephemeris measured by \citet{ray2002}.
    The time intervals in which we performed the analysis are indicated in
    the figure between vertical solid lines and labeled with the
    corresponding name.
    }
\label{fig:lc1}
\end{center}
\end{figure}

As it can be seen from the light curves of Fig.~\ref{fig:lc1},
we detected three flares of about the same intensity
(labeled H1, H2 and H3, where ``H'' is an abbreviation for \emph{High luminosity}), 
the first two occurred in correspondence to the NS
apoastron, the last while the NS was moving away from the companion.
Apoastron flares of this source have been observed also by the Burst Alert Telescope on 
board of \textsl{Swift} and reported by \citet{corbet2007}.

At MJD~53\,144 (day 1600 in the scale of Fig.~\ref{fig:lc1})
the source was so weak that it was pointless to perform any
analysis.
In the following orbital cycle, it exhibited weak and irregular luminosity
increases, none of which was intense enough to allow a dedicated analysis.
We then treated these data as a unique observation
named EI, meaning that the source had an \emph{Extended Intermediate}
luminosity level.
We divided the rest of the observations into three intervals labeled L1, L2 and
L3 (where ``L'' is an abbreviation for \emph{Low luminosity}),
during which the source exhibited less pronounced brightening. In the interval selection we
excluded the neighboring SCWs with few source counts to improve the S/N and
we separated L3 from EI since it has the appearance of a low luminosity flare occurring near 
the system periastron, such as L1, rather than an extension of EI.
We finally note that the L2 episode occurred while the NS was approaching the companion.

The H3 episode has been reported by \citet{atel}, who stated that a flare of this intensity
has never been mentioned in the literature for 1E~1145.1--6141.
But it should be noted that the source did present brightening
of the same intensity during the episodes called here H1 and H2.

To perform the timing analysis, we transformed the event arrival times
for each time interval listed in Table~\ref{tab:obs} to the Solar System
Barycenter and then to the Binary System Barycenter using the orbital parameters
found by \citet{ray2002}. We determined the pulse period
with a version of the phase-shift method described extensively in
\citet{ferrigno2006} and briefly summarized here.
Using the \textsl{ISGRI} data, which guarantee the best signal to noise ratio (S/N) and
time coverage, we find a tentative pulse period
from the Fourier analysis of the light curve. Then, using this value, we
accumulate a pulse profile in each science window and determine
the phase of the pulse maximum.
Finally, from a parabolic fit of the pulse maximum phase as a
function of time, we find the optimal pulse period and an upper limit on its first derivative.

Unless differently stated, the errors in the paper are quoted at $1\sigma$
confidence level.

%

\begin{table*}
\begin{center}
\caption{
    Summary of the \textsl{INTEGRAL} observations of 1E~1145.1-6141.
}

\begin{tabular}{ l l l c c r@{}l r@{}l r@{}l}
\hline
\hline
Name & Start [MJD] & End [MJD] &  \multicolumn{2}{c}{Exposure[ks]} & \multicolumn{4}{c}{Mean Count Rate [cts/s]} &  \multicolumn{2}{c}{L(5-50\,keV)$^*$}\\
     &             &           & \textsl{ISGRI} & \textsl{JEM-X}   & \multicolumn{2}{c}{\textsl{ISGRI}} & \multicolumn{2}{c}{\textsl{JEM-X}} & \multicolumn{2}{c}{[$\ergs$]} \\
\hline
L1   & 52\,788.8947 & 52\,791.0241 &  64.1 & 10.9 & $5.9$&$\pm0.1$   & $3.2$&$\pm0.1$  & $4.1$&$\times 10^{35}$\\
H1   & 52\,795.6284 & 52\,796.4538 &  29.2 & 13.7 & $14.4$&$\pm0.2$  & $6.2$&$\pm0.1$  & $1.1$&$\times 10^{36}$\\
L2   & 52\,814.7897 & 52\,815.5495 &  49.6 & 15.5 & $5.8$&$\pm0.2$   & $4.3$&$\pm0.5$  & $2.5$&$\times 10^{35}$\\
H2   & 52\,823.7611 & 52\,824.9043 &  39.0 &      & $14.7$&$\pm0.2$  &  &  \\
EI   & 53\,146.8561 & 53\,162.3828 & 703.5 &      & $3.53$&$\pm0.05$ &  &  \\
L3   & 53\,162.6755 & 53\,163.3679 &  42.1 &      & $7.9$&$\pm0.2$   &  & \\
H3   & 53\,165.3538 & 53\,169.1147 & 153.4 & 2.2  & $12.2$&$\pm0.1$  & $8.35$&$\pm0.03$& $1.0$&$\times 10^{36}$\\
\hline
\multicolumn{11}{l}{$^*$ assuming a distance of 8.5\,kpc.}
\label{tab:obs}
\end{tabular}
\end{center}

\end{table*}

\section{Results}
\label{sec:results}

\subsection{Timing analysis}
\label{sec:timing}

We measure the pulse period in the time intervals
reported in Table~\ref{tab:spin_period}; while during L1, L2 and L3 the statistics
is not sufficient for an independent determination.
The pulse period derivative is always too small to be directly
measured. In particular, during EI we find the $2\sigma$ upper limit $\left|\dot P\right| < 4\times10^{-9}$\,s/s.
Thus, during this orbital cycle the period variations do not exceed the long
term trend.

\begin{figure}
  \begin{center}
    \resizebox{\hsize}{!}
    {
    \includegraphics[width=8cm,angle=0]{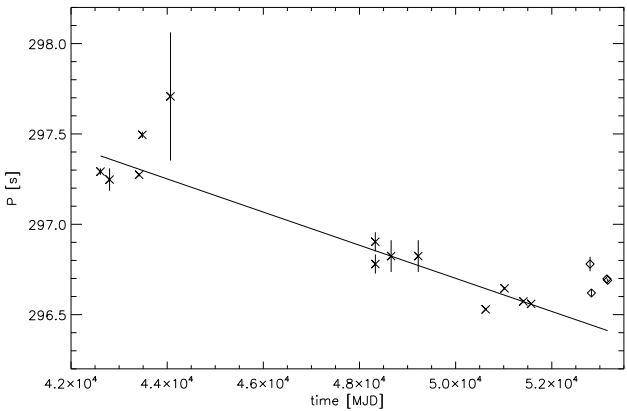}
    }
    \caption{
    Long-term evolution of the pulse period of 1E~1145.1-6141. The diamonds
    after MJD~52\,000 represent the results of our analysis. The crosses are the previous determinations,
    and the solid line is the spin-up trend derived by \citet{ray2002}.
    }
\label{fig:time_series}
\end{center}
\end{figure}

We plot in Fig.~\ref{fig:time_series} our results together with 
the period and the spin-up trend reported by \citet{ray2002}.
We note that the \textsl{INTEGRAL} values
are significantly higher than the extrapolation of
the secular evolution measured by \citet{ray2002}.
To verify the occurrence of a torque reversal, we computed the average value of our period determinations ($296.695\pm0.002$\,s)
and of the last four \textsl{RXTE} results ($296.572\pm0.001$\,s): the difference
of $0.123\pm0.003$\,s in five years evidences an average spin-down of $\sim7\times10^{-10}$\,s/s in
contrast to the previous spin-up trend of $-1 \times 10^{-9}$\,s/s. The source has then
begun to reduce its average angular speed on a time scale of years even though fluctuations from the trend
are present. In fact,
at MJD 52824.0 (H2) we found a period of $(296.62\pm0.02)$\,s,
compatible with the \textsl{RXTE} determinations, but at more than $3\sigma$ from the other values we measured.
If the trend proposed by \citet{ray2002} is correct a torque reversal
would have taken place between \textsl{RXTE} and \textsl{INTEGRAL} observations around MJD~52\,000; but we cannot exclude that the source was spinning down before MJD~45\,000, spinning up until MJD~50\,500 and then spinning up again. Unfortunately, the
poor determination of the pulse period by \textsl{EINSTEIN} at MJD~44\,000 does not allow to verify this second hypothesis.

\begin{figure}
  \begin{center}
    \resizebox{\hsize}{!}{\includegraphics[angle=0]{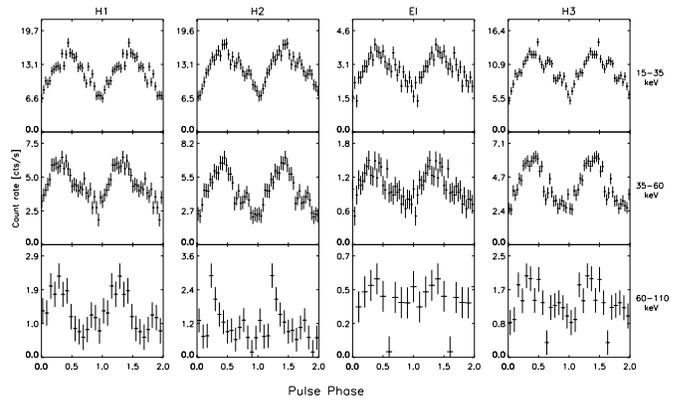}}
    \caption{Background subtracted pulse profiles of 1E~1145.1-6141 for different
    periods in three energy bands.
    }
\label{fig:pulses}
\end{center}
\end{figure}

Using the values of Table~\ref{tab:spin_period}, we produced the
pulse profiles of the \textsl{ISGRI} data in the energy bands 15--35,
35--60, and 60--110\,keV setting arbitrarily the zero phase at the pulse
minimum in the lowest energy range. The pulse profiles (Fig.~\ref{fig:pulses})
show a similar shape over all the observing periods.
They are dominated by a single broad peak which moves to earlier phase for
increasing energy, and present also a hint for a second peak at phase $\sim0.7$,
which is reminiscent of the notch between the pulse maximum and minimum
reported by \citet{grebenev1992} and \citet{ray2002}.
However, some differences arise at
higher energy where the main peak has an increasing sharpness, 
if the time intervals are put in the sequence EI,
H3, H1, and H2.

\begin{table*}
\caption{
Pulse periods ($P$) and the upper limits on its first derivatives ($\dot P$).
}
\begin{center}
\begin{tabular}{ l l r@{}l c}
\hline
\hline
name & reference time$^{*}$ [MJD] & \multicolumn{2}{c}{$P$ [s]} & $\left|\dot P\right|^{**}$ [s/s]  \\
\hline
H1  &   52\,796.00566  &    $296.78 $&$ \pm   0.04$  &$ <7 \times 10^{-6}$\\
H2  &   52\,824.25522  &    $296.62 $&$ \pm   0.02$  &$ <6.5 \times 10^{-7}$\\
EI  &   53\,154.80607  &    $296.698 $&$\pm   0.003$ &$ <4 \times 10^{-9}$\\
H3  &   53\,167.38765  &    $296.69 $&$\pm   0.01$ &$ <1.8 \times 10^{-7}$ \\
\hline
\multicolumn{5}{l}{$^{*}$At the reference time the pulse profile has a minimum in the 15--35\,keV energy range.}\\
\multicolumn{5}{l}{$^{**}$Upper limits are given at the $2\sigma$ confidence level.}
\end{tabular}
\end{center}
\label{tab:spin_period}
\end{table*}

\begin{figure}
  \begin{center}
    \resizebox{\hsize}{!}{\includegraphics[angle=0]{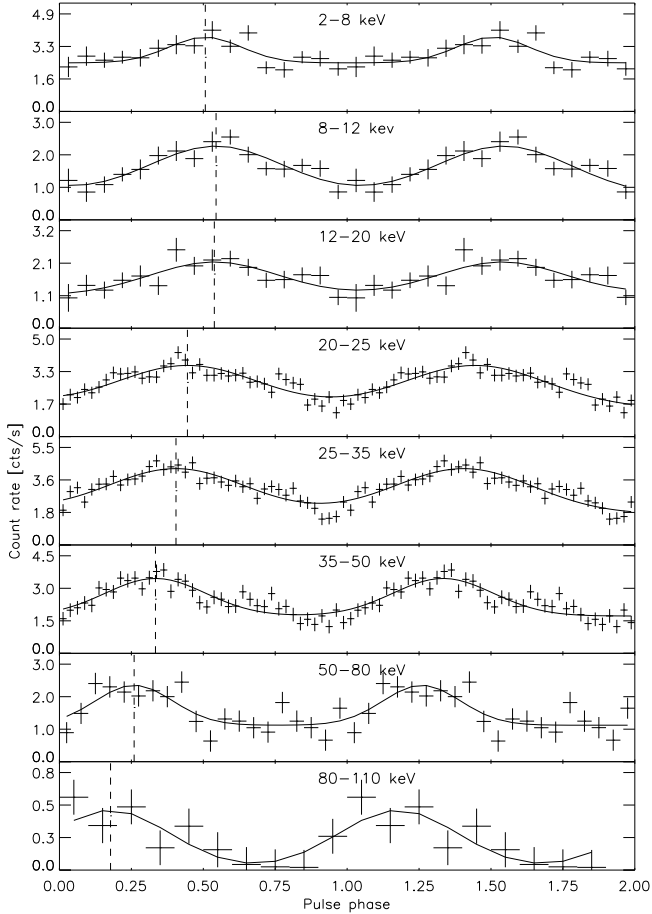}}
    \caption{Background subtracted pulse profiles of 1E~1145.1-6141 during H1. Below 20\,keV the
     data are from \jemx, above from \isgri. In each panel, the solid line is
     the best fit pulse profile obtained adding a Gaussian
     to a constant; the vertical dashed line is the position of the Gaussian
     centroid.
    }
\label{fig:pulses_h1}
\end{center}
\end{figure}

To investigate the evolution of the pulse profile with energy, we focus on
the H1 time interval when \jemx exposure is long enough to accumulate pulse
profiles with an acceptable S/N below 20\,keV.
The pulse profiles are modeled with a Gaussian plus a constant (reduced
$\chi^2$ between 0.8 and 1.4 for 35 degrees of freedom) as shown in
Fig.~\ref{fig:pulses_h1}.
We find that the width of the Gaussian remains constant with an average value
of $0.21\pm0.05$ while its centroid shifts at earlier phase for increasing energy:
between 2 and 8\,keV the
pulse profile peaks at pulse phase $0.5\pm0.2$, between 20 and 30\,keV
at pulse phase $0.4\pm0.1$, and
above 80\,keV at pulse phase $0.2\pm0.2$.
The linear fit of the phase of the centroid position $x$ as function
of energy $E$ gives:
\begin{equation}
x = (0.55\pm0.10) - (4.4 \pm 2.4)\times10^{-3} \frac{E}{\mathrm{[keV]}}
\label{eq:trend}
\end{equation}
Repeating the same procedure for the intervals H2, EI and H3 for the
\textsl{ISGRI} data alone, due to the bad quality or lack of \textsl{JEM-X} data,
we find that the relation of equation~(\ref{eq:trend}) is always verified.
However, the limited energy range prevents an independent
determination of the centroid evolution with energy.

\subsection{Spectral analysis}
In each observing period we perform broad band phase averaged spectral
analysis using \textsl{ISGRI} data and, whenever available, \textsl{JEM--X} data.
We investigate several spectral models and eventually find satisfactory fits
for cut-off power law ($E^{-\gamma}\exp\left[-E/E_f\right]$) plus
photoelectric absorption, implemented as \texttt{PHABS} in \texttt{XSPEC} using the cross sections by \citet{phabs1992},
and bremsstrahlung also attenuated by absorption.
But a discrimination between these models on a statistical base
is not possible because all fits give acceptable $\chi^2$ values (see Table~\ref{tab:fit_res}).

The cut-off power law model gives best fit values with 
large uncertainties in most of the observing periods probably
due to a strong correlation between the spectral index and the other
parameters. It is then difficult to find any significant trend of
the parameters with luminosity or time; nonetheless some conclusions can
be drawn: the lowest value of $N_\mathrm{H}$ ($7.5\pm2.5\times10^{22}\,\mathrm{cm}^{-2}$) is
significantly higher than both the Galactic column
\citep[$N_\mathrm{H}=1.4\times10^{22}\,\mathrm{cm^{-2}}$;
][]{dickey1990} and the intrinsic column measured by
\textsl{EINSTEIN} \citep[$\left( 3 \pm 2\right) \times
10^{22}\,\mathrm{cm}^{-2}$, ][]{lamb1980}; the measured spectral index of
the power law (1--1.5) is typical of several
HMXBs \citep{coburn2002}.

\begin{figure}
  \begin{center}
  \resizebox{\hsize}{!}{\includegraphics[angle=0]{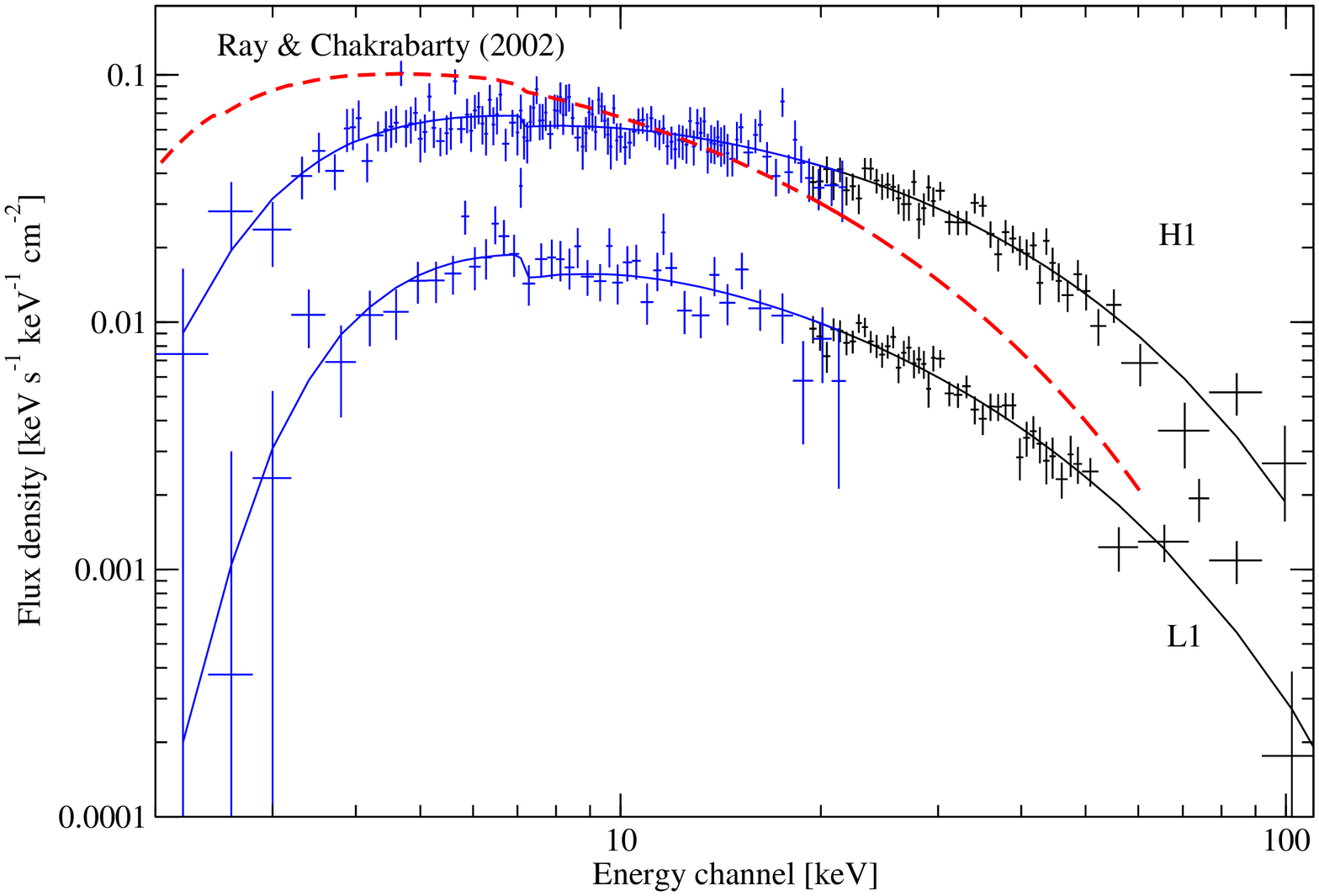}}
    \caption{Broad band unfolded energy spectra for the high state H1 and the low state L1.
    Below $\sim$20\,keV
    data are from \textsl{JEM-X}, above this value from \textsl{ISGRI}. The model in both cases is
    absorbed bremsstrahlung. The dashed line is the model used by
    \citet{ray2002} with arbitrary normalization.}
\label{fig:spec}
\end{center}
\end{figure}

In Fig.~\ref{fig:spec}, we show the spectra in the low and high
luminosity states L1 and H1 together with the high energy cut-off model used by \citet{ray2002}
($E^{-\gamma}\exp\left[-(E-E_c)/E_f\right]$ for $E>E_c$ and $E^{-\gamma}$ for $E<E_c$).
We note that in our data the source emission is considerably harder,
as confirmed by the higher value of the folding energy
$E_f=28^{+5}_{-3}$\,keV when the spectra during H1 and H3 are
described by the same model of \citet{ray2002}.
Fitting the spectra of the other observing periods with this function, we note
unreasonably large uncertainties on the parameters together with a $\chi^2$ not better than
the values of Table~\ref{tab:fit_res}. It is then evident that 
the relatively low S/N of the \textsl{INTEGRAL} observations of this source
prevents the use of this phenomenological model, which has a redundant number of free parameters.
We finally note that, since the source brightness in our observations is not greater than in the
past, we can exclude that this evolution is luminosity related.

\begin{figure}
  \begin{center}
    \resizebox{\hsize}{!}{\includegraphics[angle=0]{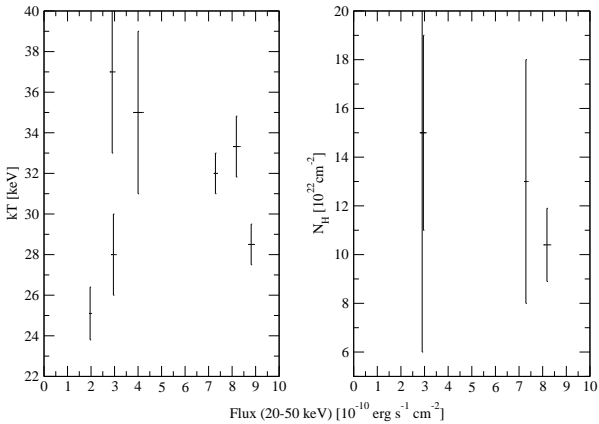}}
    \caption{Best fit parameters of the absorbed bremsstrahlung model as
     function of the source luminosity.
     In the left panel we plot the plasma temperature and in the right panel
     the $N_\mathrm{H}$ of the intrinsic photoelectric absorber.
     Errors are at 90\% c.l. for one parameter of interest.
    }
\label{fig:par}
\end{center}
\end{figure}

From the best fit values of the bremsstrahlung model plotted in
Fig.~\ref{fig:par}, we note that the plasma temperature is variable
even if not correlated with the source luminosity.
Instead, $N_\mathrm{H}$ is always
higher than the Galactic column and does not present any significant
trend. We also verify that fixing the absorbing column at its
average value, the determinations of the plasma temperature do not
vary significantly.

\subsection{Phase resolved spectral analysis}
During the time interval H1 the statistics allows to perform broad band
phase resolved spectroscopy combining data from both \textsl{ISGRI}
and \textsl{JEM-X}. In the other time intervals the source is either too weak or
too off-axis for \textsl{JEM-X} to collect enough counts, and then
to significantly constrain the continuum parameters.

\begin{figure}
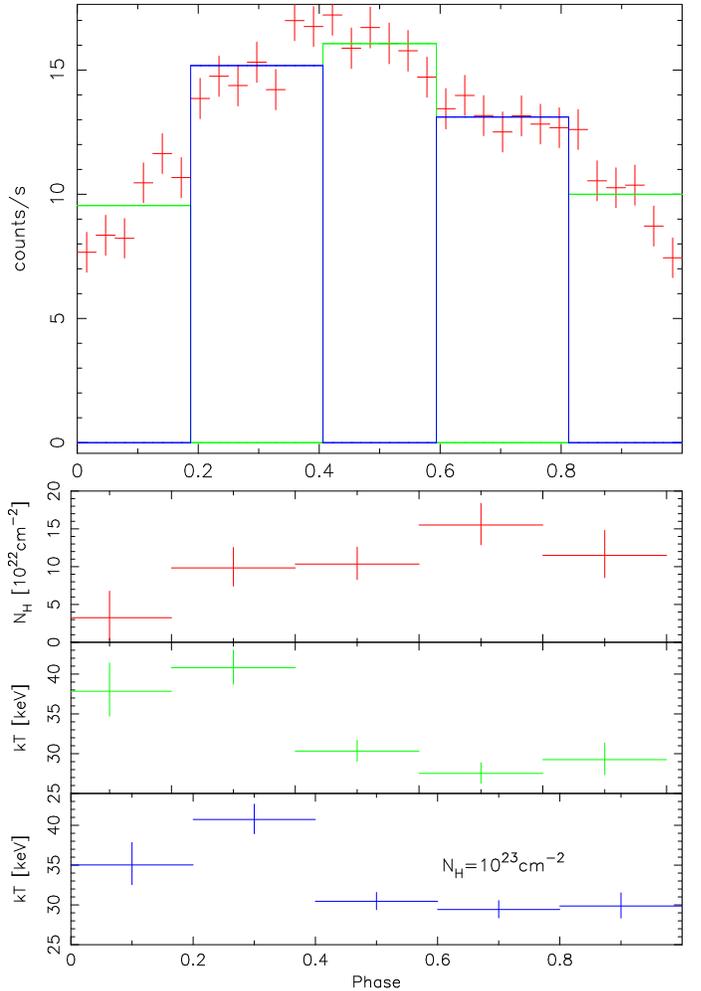

  \begin{center}
    \resizebox{\hsize}{!}{\includegraphics[angle=270]{phases.ps}}
    \resizebox{\hsize}{!}{\includegraphics[angle=270]{parametri.ps}}
    \caption{Phase resolved spectral analysis results. Upper panel: the pulse profile during H1 in the 20--60\,keV
      energy range with the adopted phase intervals. Lower panel: the best fit parameters of the adopted model
      (photo-electric absorbed bremsstrahlung) as a function of the pulse phase
      during H1. From top to bottom we plot the hydrogen column density $N_\mathrm{H}$, the plasma
      temperature $kT$, when $N_\mathrm{H}$ is a free parameter and when $N_\mathrm{H}$ is held fixed at
      $10^{23}\,\mathrm{cm^{-2}}$.
      }
\label{fig:phase}
\end{center}
\end{figure}

In this analysis we adopt the absorbed bremsstrahlung model, since it gives a good
description of the source emission with the lowest number of parameters.
We divide the pulse in five equally spaced phase intervals and
fit the spectra obtaining a reduced $\chi^2\sim1$ for $\sim$100 d.o.f. and the
the best fit parameters plotted in Fig.~\ref{fig:phase}.
The value of $N_\mathrm{H}$ is approximately constant
($\sim10^{23}\,\mathrm{cm}^{-2}$),
while the plasma temperature is higher at pulse phase 0--0.4
(ascending edge of the pulse) than at pulse phase 0.4--1 (descending edge).
Averaging the fitted values,
we find $kT_{0-0.4}=(40\pm2)$\,keV and $kT_{0.4-1}=(29\pm1)$\,keV
corresponding to a variation of the temperature with a significance of $5\sigma$.
Fixing the value of $N_\mathrm{H}$ at $10^{23}\,\mathrm{cm}^{-2}$ (the average over the spin period)
we obtain temperature determinations compatible to the previous case,
and a slight improvement in the significance of the variation ($5.3\sigma$).
The higher temperature in the rising part of the main
peak explains naturally its shift
with energy since at increasing energy the hotter emission is prominent.

The phase resolved spectral analysis in the other time intervals does not
give a statistically significant variation of the plasma temperature.

\begin{table*}
\begin{center}
\caption{
	Summary of the results of spectral fits with two models:
        cut-off power law and thermal bremsstrahlung, both modified by
        photoelectric absorption.
	The fluxes are computed directly from the source counts and are thus independent from the
	adopted absorption; for comparison the flux of the Crab in the
	20-50\,keV energy band is $10^{-8} \mathrm{erg\,s^{-1}\,cm^{-2}}$.
    The errors are at 90\% confidence level for one parameter of
    interest ($\chi^2_{\mathrm{min}}+2.7$).
}
\begin{tabular}{l c c c c c c c}
\hline
\hline
parameter                           & L1           & H1            & L2          & H2           & EI           & L3          & H3 \\
\hline
\multicolumn{3}{l}{Cut-off power law} & & & & & \\
$N_H\; [10^{22}\, \mathrm{cm^{-2}}]$& $20\pm8$        & $7.0\pm2.5$   &$19_{-11}^{+15}$ & N/A            &      N/A              & N/A            & $12\pm7$ \\
Photon index                        & $1.5\pm0.4$     & $1.1\pm0.1$ & $1.5\pm0.5$     & $1.4\pm0.2$    & $1.8\pm0.3$      & $1.1\pm0.8$    & $1.2\pm0.2$ \\
$E_f$ [keV]                         & $31_{-8}^{+16}$ & $25.5\pm2.5$  & $39_{-13}^{+50}$& $28^{+7}_{-5}$ & $35_{-10}^{+18}$ & $25_{-9}^{+31}$ &  $29^{+5}_{-4}$\\
$\chi^2$/d.o.f.                     & 218/197         & 232/220       & 240/205         & 36/43          &   64/51            & 74/65          & 197/195       \\
\hline
\multicolumn{3}{l}{Bremsstrahlung} & & & & & \\
$N_H\; [10^{22}\, \mathrm{cm^{-2}}]$& $15\pm4   $  & $10.4\pm1.5$  & $15\pm9$    & N/A          & N/A          & N/A         & $13\pm5$ \\
$kT$ [keV]                          & $28\pm2 $    & $33.3\pm1.5$  & $37\pm4$    & $28.5\pm1.0$ & $25.1\pm1.3$ & $35\pm4$    & $32\pm1$\\
$\chi^2$/d.o.f.                     & 220/198      & 240/221       & 240/206     & 36/44        & 68/52        & 62/44       & 198/196       \\
\hline
flux(5--15\,keV)$^*$                   & $2.9\pm0.2$  & $7.47\pm0.15$ & $2.3\pm0.5$ & N/A          & N/A          & N/A         & $6.6\pm0.3$   \\
flux(20--50\,keV)$^*$                  & $2.9\pm0.1$& $8.2\pm0.1$ & $2.9\pm0.1$ & $8.8\pm0.1$  & $1.96\pm0.04$& $4.0\pm0.2$ & $7.29\pm0.07$ \\
\hline
\multicolumn{8}{l}{$^*$ in unity of $10^{-10} \mathrm{erg\,s^{-1}\,cm^{-2}}$.}
\end{tabular}
\label{tab:fit_res}
\end{center}
\end{table*}

\section{Discussion}
\label{sec:discussion}

We obtain a good description of the source spectral energy distribution
using the absorbed bremsstrahlung model; this does not mean, alone, that
this is the real emission mechanism. 
To make a consistency check, we compute from the expression of 
the thermal bremsstrahlung emissivity, equation (87) of \citet{longair1999},
the quantity $n_e^2V$, where $n_e$ is the electron density and $V$ is the volume of the emitting region
for a pure hydrogen plasma emitting at the observed
temperature with a power compatible to the luminosity of the source.
Assuming for the accretion column 
a radius of 100\,m and estimating the mass accretion rate from the X-ray luminosity, it is possible
to derive an order of magnitude estimation for the electron density which implies a Compton optical
depth of order unity both in the vertical and in the horizontal
directions of the accretion column. Therefore, it is not possible to
ensure that Compton scattering is negligible, but, on the other
hand, it is not clear how much it distorts the underlying spectral energy
distribution.

\citet{becker2007} recently showed that the observed spectrum of XRBPs is
well described by thermal and bulk Comptonization of
the optically thin bremsstrahlung and cyclotron emission from the accretion column.
Using the measured luminosity and distance of the source, we tested their model on the
spectrum with best S/N, i.e. during H1. The fit was very good evidencing that all the emission
comes from re-processing of an underlying bremsstrahlung with an electron temperature of
$\sim13$\,keV without any need for intrinsic absorption. Unfortunately, it was not possible to constrain 
all the model parameters, especially the magnetic field,
due to the absence of absorption features.
However, we note that the spectral energy distribution
is very similar to absorbed bremsstrahlung and then, in data with relatively low
S/N, the two models are statistically equivalent.




The values we find for the plasma temperature 
(higher than for the Becker and Wolff model)
and for the absorbing column depth should then be regarded as a useful parametrization of the
spectral energy distribution, rather than a measurement of the
plasma characteristics. The strong absorption could be due to an emission
turn-off at low energy, but could also be partially explained by the presence
of optically thick gas in the stellar wind e.g. a bow shock
formed by the X-ray pulsar radiation \citep{blondin1990}.
Unfortunately, only observations with a better spectral resolutions
below $\sim10$\,keV would detect the absorption edges
and then constrain significantly the column value.


More insight into the structure of the emission region can be achieved from the
pulse phase resolved spectroscopy.
The constant $N_\mathrm{H}$ at any phase indicates that the
low energy part of the spectrum is governed by a process isotropically
distributed around the NS or located far from it, while the bimodal temperature distribution
and the presence of
a notch in some pulse profiles (Fig.~\ref{fig:pulses}) suggest
the presence of two components at different temperatures in the magnetosphere
of 1E~1145.1-6141.
If their relative intensity is variable with time, the
different values of the plasma temperature in the phase averaged spectra between 25 and 40\,keV
and the time modulation of the pulse shape would be naturally justified.
Indeed, all the analyzed spectra can be fitted with a linear combination of
two bremsstrahlung model with fixed temperatures of 19 and 40\,keV.

During three flares, the source reached 80--90\,mCrab in the 20--50\,keV
energy band. During three other periods it reached 30--40\,mCrab,
but for one binary orbit (EI) it presented only
irregular and weak luminosity variation with an average flux of 20\,mCrab.
Two of the most intense flares occurred at the apoastron,
the other at about two thirds of the maximum distance.
At the periastron the NS reached at most one third of the maximum luminosity.

This behavior rules out the hypothesis of spherically symmetric wind, and of a
combination of disk and wind proposed to explain the periodic flares
in IGR~J11215-5952 \citep{sidoli2007} or GX~301-2 \citep{pravdo2001}. 
A complex wind environment with clumps or filaments of
higher density plasma is a more natural explanation of its irregular 
luminosity increases, but our measurements do not allow a good spectral constraint of the
absorbing column, thus this hypothesis cannot be tested spectroscopically.

A clumpy stellar wind is invoked also to explain the light curves of the super-giant fast X-ray transients \citep{jean2005,leyder2007}
characterized by short irregular flares lasting from a few ks to a fraction of a day.
The NS in these systems is thought to orbit at $\sim10$ stellar radii while in persistent systems
the separation is supposed to be smaller \citep{walter2007}.
We can estimate for 1E~1145.1-6141, on the basis of the 
spectral type, a mass of $\sim10\,M_\odot$ and a stellar radius 
of $\sim5\,R_\odot$ \citep{habets1981}. From the measured orbital parameters, 
we derive that the NS lies at about 10 stellar radii from the companion, with a distance modulation
of 20\% due to the orbit eccentricity.

Although it is a persistent system, it has some characteristics similar to the SFXT:
in particular the component separation and the flaring activity.
However, compared to these objects, the flares are longer (from a significant fraction of a day to about four days) and have
a smaller relative (4--5) intensity with respect to the persistent flux level, which, in turn, is
two orders of magnitude larger. Supposing that the mechanism to capture the 
plasma has similar efficiency in all systems, we infer that the wind is on average denser, 
while the clumps are less compact than in SFXT,
making the system an intermediate case between transient and persistent wind-fed HMXBs with super-giant companion.

Unfortunately, in our case, an estimate of the mass spectrum of the clumps
from the flare duration is not straightforward since the torque history
is characterized by local variations, and then the formation
of a temporary accretion disk is probable, which prevents the plasma from directly falling onto the NS surface.
Wind clumpyness appears a common feature in binary early type stellar wind suggesting further investigations
to clarify whether this can be due to the interaction with the NS 
or it is an intrinsic characteristic of the wind.

\section{Conclusions}
\label{sec:conclusions}

We confirm that 1E~1145.1-6141 is a persistent low luminosity binary system
but we find interesting peculiarities not previously noticed.

The source was always detectable, when in the instrument field of view, and
showed a complex light curve which strongly suggests a clumpy stellar wind.

The timing analysis shows that the pulse period of the source is not constant
(at 98\% confidence level). These variations of the torque are expected in wind-fed systems,
in addition we find that the source reversed its long term torque. In fact, the averaged of the measured pulse periods
is higher than the average of the \textsl{RXTE} observation (June 1997--January 2000) while \citet{ray2002}
found a previous spin-up trend. Unfortunately, the lack of continuous monitoring prevents to 
study in more detail the torque history and then the accretion model.

The absorbed bremsstrahlung model allows an
incisive parametrization of the source spectral characteristics, even though it cannot
be regarded as a coherent physical model, 
since the Compton optical depth of the accretion column is of order of unity.
We find that the plasma temperature
varies with no correlation to the source luminosity while the absorbing column
remains approximately constant.
A comparison with the model adopted in the previous \textsl{RXTE} observations
\citep{ray2002} indicates a significant hardening of the spectrum
which cannot be explained by a different luminosity level.

Phase resolved spectroscopy
shows that the absorbing column depth is approximately constant, and
the plasma temperature in the first half of the pulse profile is $(11\pm2)$\,keV higher than in
the second half. This justifies the pulse maximum shift by $\sim0.4$ phase units between 20 and 100\,keV.
To describe the emission of 1E~1145.1-6141, we then propose the presence
of two spectral components at different temperatures
and, possibly, a layer of absorbing material
surrounding isotropically the NS or located far from it.
The variation of the components' relative intensity with time would also explain
the oscillations of the phase averaged temperature that we observe.

\section*{Acknowledgments}
The observational data used in this communication were
collected by INTEGRAL, an ESA science mission for
X-ray and Gamma-ray astronomy. The work was supported
by the Italian Space Agency (ASI) under contract no.~I/R/023/05 and
by the German Space Agency (DLR) under
contract nos.~50~OG~9601 and 50~OG~0501.
C.F. would like to thank Rosario Iaria and the anonymous referee for their 
useful comments and suggestions.

\bibliographystyle{aa}
\bibliography{pulsars}

\end{document}

%% file: mymacros.tex
\newcommand{\isgri}{\textsl{ISGRI}\;}
\newcommand{\jemx}{\textsl{JEM-X}\;}

\newcommand{\ergs}{\ensuremath{ \mathrm{erg\,s}^{-1} }}

\newcommand{\gra}{\raisebox{1ex}{\scriptsize o}}

\def\fun#1#2{\lower3.6pt\vbox{\baselineskip0pt\lineskip.9pt
  \ialign{$\mathsurround=0pt#1\hfil##\hfil$\crcr#2\crcr\sim\crcr}}}

{\catcode`\|=\active
  \gdef\Braket#1{\left<\mathcode`\|"8000\let|\bravert {#1}\right>}}
\def\bravert{\egroup\,\vrule\,\bgroup}

\def\gsim{\lower.5ex\hbox{\gtsima}}
\newfont{\mc}{cmcsc10 scaled\magstep2}
\newfont{\cmc}{cmcsc10 scaled\magstep1}

\newcommand{\bgi}{\begin{itemize}}
\newcommand{\edi}{\end{itemize}}

\newcommand{\be}{\begin{equation}}
\newcommand{\ee}{\end{equation}}

\newcommand{\bea}{\begin{eqnarray}}                  %
\newcommand{\eea}{\end{eqnarray}}                    %

\newcommand{\beaa}{\begin{eqnarray*}}                %
\newcommand{\eeaa}{\end{eqnarray*}}                  %

\newcommand{\bgd}{\begin{description}}
\newcommand{\edd}{\end{description}}

\newcommand{\bgf}{\begin{figure}}
\newcommand{\edf}{\end{figure}}

\newcommand{\bgc}{\begin{center}}
\newcommand{\edc}{\end{center}}

\newcommand{\bgt}{\begin{tabular}}
\newcommand{\edt}{\end{tabular}}

\newcommand{\bge}{\begin{enumerate}}
\newcommand{\ede}{\end{enumerate}}

\DeclareRobustCommand{\ion}[2]{%
 \relax\ifmmode
 \ifx\testbx\f@series
 {\mathbf{#1\,\mathsc{#2}}}\else
 {\mathrm{#1\,\mathsc{#2}}}\fi
 \else\textup{#1\,{\mdseries\textsc{#2}}}%
 \fi}

%% file: aasm.tex
\def\jnl@style{\it}
\def\aaref@jnl#1{{\jnl@style#1}}

\def\aaref@jnl#1{{\jnl@style#1}}

\def\aj{\aaref@jnl{AJ}}                   
\def\araa{\aaref@jnl{ARA\&A}}             
\def\apj{\aaref@jnl{ApJ}}                 
\def\apjl{\aaref@jnl{ApJ}}                
\def\apjs{\aaref@jnl{ApJS}}               
\def\ao{\aaref@jnl{Appl.~Opt.}}           
\def\apss{\aaref@jnl{Ap\&SS}}             
\def\aap{\aaref@jnl{A\&A}}                
\def\aapr{\aaref@jnl{A\&A~Rev.}}          
\def\aaps{\aaref@jnl{A\&AS}}              
\def\azh{\aaref@jnl{AZh}}                 
\def\baas{\aaref@jnl{BAAS}}               
\def\jrasc{\aaref@jnl{JRASC}}             
\def\memras{\aaref@jnl{MmRAS}}            
\def\mnras{\aaref@jnl{MNRAS}}             
\def\pra{\aaref@jnl{Phys.~Rev.~A}}        
\def\prb{\aaref@jnl{Phys.~Rev.~B}}        
\def\prc{\aaref@jnl{Phys.~Rev.~C}}        
\def\prd{\aaref@jnl{Phys.~Rev.~D}}        
\def\pre{\aaref@jnl{Phys.~Rev.~E}}        
\def\prl{\aaref@jnl{Phys.~Rev.~Lett.}}    
\def\pasp{\aaref@jnl{PASP}}               
\def\pasj{\aaref@jnl{PASJ}}               
\def\qjras{\aaref@jnl{QJRAS}}             
\def\skytel{\aaref@jnl{S\&T}}             
\def\solphys{\aaref@jnl{Sol.~Phys.}}      
\def\sovast{\aaref@jnl{Soviet~Ast.}}      
\def\ssr{\aaref@jnl{Space~Sci.~Rev.}}     
\def\zap{\aaref@jnl{ZAp}}                 
\def\nat{\aaref@jnl{Nature}}              
\def\iaucirc{\aaref@jnl{IAU~Circ.}}       
\def\aplett{\aaref@jnl{Astrophys.~Lett.}} 
\def\apspr{\aaref@jnl{Astrophys.~Space~Phys.~Res.}}
\def\bain{\aaref@jnl{Bull.~Astron.~Inst.~Netherlands}} 
\def\fcp{\aaref@jnl{Fund.~Cosmic~Phys.}}  
\def\gca{\aaref@jnl{Geochim.~Cosmochim.~Acta}}   
\def\grl{\aaref@jnl{Geophys.~Res.~Lett.}} 
\def\jcp{\aaref@jnl{J.~Chem.~Phys.}}      
\def\jgr{\aaref@jnl{J.~Geophys.~Res.}}    
\def\jqsrt{\aaref@jnl{J.~Quant.~Spec.~Radiat.~Transf.}}
\def\memsai{\aaref@jnl{Mem.~Soc.~Astron.~Italiana}}
\def\nphysa{\aaref@jnl{Nucl.~Phys.~A}}   
\def\physrep{\aaref@jnl{Phys.~Rep.}}   
\def\physscr{\aaref@jnl{Phys.~Scr}}   
\def\planss{\aaref@jnl{Planet.~Space~Sci.}}   
\def\procspie{\aaref@jnl{Proc.~SPIE}}   

%% file: ferrigno_1E1145.bbl
\begin{thebibliography}{39}
\expandafter\ifx\csname natexlab\endcsname\relax\def\natexlab#1{#1}\fi

\bibitem[{{Balucinska-Church} \& {McCammon}(1992)}]{phabs1992}
{Balucinska-Church}, M. \& {McCammon}, D. 1992, \apj, 400, 699

\bibitem[{{Baykal}(1997)}]{baykal1997}
{Baykal}, A. 1997, \aap, 319, 515

\bibitem[{{Becker} \& {Wolff}(2007)}]{becker2007}
{Becker}, P.~A. \& {Wolff}, M.~T. 2007, \apj, 654, 435

\bibitem[{{Bildsten} {et~al.}(1997){Bildsten}, {Chakrabarty}, {Chiu}, {Finger},
  {Koh}, {Nelson}, {Prince}, {Rubin}, {Scott}, {Stollberg}, {Vaughan},
  {Wilson}, \& {Wilson}}]{bildsten1997}
{Bildsten}, L., {Chakrabarty}, D., {Chiu}, J., {et~al.} 1997, \apjs, 113, 367

\bibitem[{{Blondin} {et~al.}(1990){Blondin}, {Kallman}, {Fryxell}, \&
  {Taam}}]{blondin1990}
{Blondin}, J.~M., {Kallman}, T.~R., {Fryxell}, B.~A., \& {Taam}, R.~E. 1990,
  \apj, 356, 591

\bibitem[{{Bodaghee} {et~al.}(2004){Bodaghee}, {Mowlavi}, \& {Ballet}}]{atel}
{Bodaghee}, A., {Mowlavi}, N., \& {Ballet}, J. 2004, The Astronomer's Telegram,
  290, 1

\bibitem[{{Coburn} {et~al.}(2002){Coburn}, {Heindl}, {Rothschild}, {Gruber},
  {Kreykenbohm}, {Wilms}, {Kretschmar}, \& {Staubert}}]{coburn2002}
{Coburn}, W., {Heindl}, W.~A., {Rothschild}, R.~E., {et~al.} 2002, \apj, 580,
  394

\bibitem[{{Corbet} {et~al.}(2007){Corbet}, {Markwardt}, {Barbier}, {Barthelmy},
  {Cummings}, {Gehrels}, {Krimm}, {Palmer}, {Sakamoto}, {Sato}, \&
  {Tueller}}]{corbet2007}
{Corbet}, R., {Markwardt}, C., {Barbier}, L., {et~al.} 2007, ArXiv e-prints,
  astro-ph/0703274

\bibitem[{{Davidson} \& {Ostriker}(1973)}]{Davids73}
{Davidson}, K. \& {Ostriker}, J.~P. 1973, \apj, 179, 585

\bibitem[{{Densham} \& {Charles}(1982)}]{densham1982}
{Densham}, R.~H. \& {Charles}, P.~A. 1982, \mnras, 201, 171

\bibitem[{{di Salvo} {et~al.}(2004){di Salvo}, {Santangelo}, \&
  {Segreto}}]{disalvo2004}
{di Salvo}, T., {Santangelo}, A., \& {Segreto}, A. 2004, Nuclear Physics B
  Proceedings Supplements, 132, 446

\bibitem[{{Dickey} \& {Lockman}(1990)}]{dickey1990}
{Dickey}, J.~M. \& {Lockman}, F.~J. 1990, \araa, 28, 215

\bibitem[{{Ferrigno} {et~al.}(2007){Ferrigno}, {Segreto}, {Santangelo},
  {Wilms}, {Kreykenbohm}, {Denis}, \& {Staubert}}]{ferrigno2006}
{Ferrigno}, C., {Segreto}, A., {Santangelo}, A., {et~al.} 2007, \aap, 462, 995

\bibitem[{{Giacconi} {et~al.}(1971){Giacconi}, {Gursky}, {Kellogg}, {Schreier},
  \& {Tananbaum}}]{giac71}
{Giacconi}, R., {Gursky}, H., {Kellogg}, E., {Schreier}, E., \& {Tananbaum}, H.
  1971, \apjl, 167, L67

\bibitem[{{Grebenev} {et~al.}(1992){Grebenev}, {Pavlinsk}, \&
  {Syunyaev}}]{grebenev1992}
{Grebenev}, S.~A., {Pavlinsk}, M.~N., \& {Syunyaev}, R.~A. 1992, Soviet
  Astronomy Letters, 18, 228

\bibitem[{{Habets} \& {Heintze}(1981)}]{habets1981}
{Habets}, G.~M.~H.~J. \& {Heintze}, J.~R.~W. 1981, \aaps, 46, 193

\bibitem[{{Hutchings} {et~al.}(1981){Hutchings}, {Crampton}, \&
  {Cowley}}]{hutchings1981}
{Hutchings}, J.~B., {Crampton}, D., \& {Cowley}, A.~P. 1981, \aj, 86, 871

\bibitem[{{{\.I}nam} \& {Baykal}(2000)}]{turchi2000}
{{\.I}nam}, S.~{\c C}. \& {Baykal}, A. 2000, \aap, 353, 617

\bibitem[{{in't Zand}(2005)}]{jean2005}
{in't Zand}, J.~J.~M. 2005, \aap, 441, L1

\bibitem[{{Koh} {et~al.}(1997){Koh}, {Bildsten}, {Chakrabarty}, {Nelson},
  {Prince}, {Vaughan}, {Finger}, {Wilson}, \& {Rubin}}]{koh1997}
{Koh}, D.~T., {Bildsten}, L., {Chakrabarty}, D., {et~al.} 1997, \apj, 479, 933

\bibitem[{{Labanti} {et~al.}(2003){Labanti}, {Di Cocco}, {Ferro}, {Gianotti},
  {Mauri}, {Rossi}, {Stephen}, {Traci}, \& {Trifoglio}}]{picsit}
{Labanti}, C., {Di Cocco}, G., {Ferro}, G., {et~al.} 2003, \aap, 411, L149

\bibitem[{{Lamb} {et~al.}(1980){Lamb}, {Markert}, {Hartman}, {Thompson}, \&
  {Bignami}}]{lamb1980}
{Lamb}, R.~C., {Markert}, T.~H., {Hartman}, R.~C., {Thompson}, D.~J., \&
  {Bignami}, G.~F. 1980, \apj, 239, 651

\bibitem[{{Lebrun} {et~al.}(2003){Lebrun}, {Leray}, {Lavocat}, {Cr{\' e}tolle},
  {Arqu{\` e}s}, {Blondel}, {Bonnin}, {Bou{\` e}re}, {Cara}, {Chaleil}, {Daly},
  {Desages}, {Dzitko}, {Horeau}, {Laurent}, {Limousin}, {Mathy}, {Mauguen},
  {Meignier}, {Molini{\' e}}, {Poindron}, {Rouger}, {Sauvageon}, \&
  {Tourrette}}]{isgri}
{Lebrun}, F., {Leray}, J.~P., {Lavocat}, P., {et~al.} 2003, \aap, 411, L141

\bibitem[{{Leyder} {et~al.}(2007){Leyder}, {Walter}, {Lazos}, {Masetti}, \&
  {Produit}}]{leyder2007}
{Leyder}, J.-C., {Walter}, R., {Lazos}, M., {Masetti}, N., \& {Produit}, N.
  2007, \aap, 465, L35

\bibitem[{{Longair}(1999)}]{longair1999}
{Longair}, M.~S. 1999, in LNP Vol. 520: X-Ray Spectroscopy in Astrophysics, ed.
  J.~{van Paradijs} \& J.~A.~M. {Bleeker}, pag.~24

\bibitem[{{Lund} {et~al.}(2003){Lund}, {Budtz-J{\o}rgensen}, {Westergaard},
  {Brandt}, {Rasmussen}, {Hornstrup}, {Oxborrow}, {Chenevez}, {Jensen},
  {Laursen}, {Andersen}, {Mogensen}, {Rasmussen}, {Om{\o}}, {Pedersen},
  {Polny}, {Andersson}, {Andersson}, {K{\" a}m{\" a}r{\" a}inen}, {Vilhu},
  {Huovelin}, {Maisala}, {Morawski}, {Juchnikowski}, {Costa}, {Feroci},
  {Rubini}, {Rapisarda}, {Morelli}, {Carassiti}, {Frontera}, {Pelliciari},
  {Loffredo}, {Mart{\'{\i}}nez N{\' u}{\~ n}ez}, {Reglero}, {Velasco},
  {Larsson}, {Svensson}, {Zdziarski}, {Castro-Tirado}, {Attina}, {Goria},
  {Giulianelli}, {Cordero}, {Rezazad}, {Schmidt}, {Carli}, {Gomez}, {Jensen},
  {Sarri}, {Tiemon}, {Orr}, {Much}, {Kretschmar}, \& {Schnopper}}]{jemx}
{Lund}, N., {Budtz-J{\o}rgensen}, C., {Westergaard}, N.~J., {et~al.} 2003,
  \aap, 411, L231

\bibitem[{{Negueruela} {et~al.}(2006){Negueruela}, {Smith}, {Reig}, {Chaty}, \&
  {Torrej{\'o}n}}]{negueruela2006}
{Negueruela}, I., {Smith}, D.~M., {Reig}, P., {Chaty}, S., \& {Torrej{\'o}n},
  J.~M. 2006, in ESA Special Publication, Vol. 604, The X-ray Universe 2005,
  ed. A.~{Wilson}, 165--170

\bibitem[{{Pravdo} \& {Ghosh}(2001)}]{pravdo2001}
{Pravdo}, S.~H. \& {Ghosh}, P. 2001, \apj, 554, 383

\bibitem[{{Pringle} \& {Rees}(1972)}]{Pringle72}
{Pringle}, J.~E. \& {Rees}, M.~J. 1972, \aap, 21, 1

\bibitem[{{Ray} \& {Chakrabarty}(2002)}]{ray2002}
{Ray}, P.~S. \& {Chakrabarty}, D. 2002, \apj, 581, 1293

\bibitem[{{Segreto} \& {Ferrigno}(2007)}]{segreto2007}
{Segreto}, A. \& {Ferrigno}, C. 2007, ArXiv e-prints, astrp-ph/07094132

\bibitem[{{Sguera} {et~al.}(2005){Sguera}, {Barlow}, {Bird}, {Clark}, {Dean},
  {Hill}, {Moran}, {Shaw}, {Willis}, {Bazzano}, {Ubertini}, \&
  {Malizia}}]{sguera2005}
{Sguera}, V., {Barlow}, E.~J., {Bird}, A.~J., {et~al.} 2005, \aap, 444, 221

\bibitem[{{Sguera} {et~al.}(2006){Sguera}, {Bazzano}, {Bird}, {Dean},
  {Ubertini}, {Barlow}, {Bassani}, {Clark}, {Hill}, {Malizia}, {Molina}, \&
  {Stephen}}]{sguera2006}
{Sguera}, V., {Bazzano}, A., {Bird}, A.~J., {et~al.} 2006, \apj, 646, 452

\bibitem[{{Sidoli} {et~al.}(2007){Sidoli}, {Romano}, {Mereghetti}, {Paizis},
  {Vercellone}, {Mangano}, \& {Gotz}}]{sidoli2007}
{Sidoli}, L., {Romano}, P., {Mereghetti}, S., {et~al.} 2007, ArXiv e-prints,
  astro-ph/07101175

\bibitem[{{Stevens} {et~al.}(1997){Stevens}, {Reig}, {Coe}, {Buckley},
  {Fabregat}, \& {Steele}}]{stevens1997}
{Stevens}, J.~B., {Reig}, P., {Coe}, M.~J., {et~al.} 1997, \mnras, 288, 988

\bibitem[{{Ubertini} {et~al.}(2003){Ubertini}, {Lebrun}, {Di Cocco}, {Bazzano},
  {Bird}, {Broenstad}, {Goldwurm}, {La Rosa}, {Labanti}, {Laurent}, {Mirabel},
  {Quadrini}, {Ramsey}, {Reglero}, {Sabau}, {Sacco}, {Staubert}, {Vigroux},
  {Weisskopf}, \& {Zdziarski}}]{ibis}
{Ubertini}, P., {Lebrun}, F., {Di Cocco}, G., {et~al.} 2003, \aap, 411, L131

\bibitem[{{Vedrenne} {et~al.}(2003){Vedrenne}, {Roques}, {Sch{\" o}nfelder},
  {Mandrou}, {Lichti}, {von Kienlin}, {Cordier}, {Schanne}, {Kn{\" o}dlseder},
  {Skinner}, {Jean}, {Sanchez}, {Caraveo}, {Teegarden}, {von Ballmoos},
  {Bouchet}, {Paul}, {Matteson}, {Boggs}, {Wunderer}, {Leleux},
  {Weidenspointner}, {Durouchoux}, {Diehl}, {Strong}, {Cass{\' e}}, {Clair}, \&
  {Andr{\' e}}}]{spi}
{Vedrenne}, G., {Roques}, J.-P., {Sch{\" o}nfelder}, V., {et~al.} 2003, \aap,
  411, L63

\bibitem[{{Walter} \& {Zurita Heras}(2007)}]{walter2007}
{Walter}, R. \& {Zurita Heras}, J. 2007, \aap, 476, 335

\bibitem[{{Winkler} {et~al.}(2003){Winkler}, {Courvoisier}, {Di Cocco},
  {Gehrels}, {Gim{\'e}nez}, {Grebenev}, {Hermsen}, {Mas-Hesse}, {Lebrun},
  {Lund}, {Palumbo}, {Paul}, {Roques}, {Schnopper}, {Sch{\"o}nfelder},
  {Sunyaev}, {Teegarden}, {Ubertini}, {Vedrenne}, \& {Dean}}]{integral}
{Winkler}, C., {Courvoisier}, T.~J.-L., {Di Cocco}, G., {et~al.} 2003, \aap,
  411, L1

\end{thebibliography}
